\documentclass[sigconf]{acmart}

\usepackage{booktabs} % For formal tables
\renewcommand\footnotetextcopyrightpermission[1]{} % removes footnote with conference information in first column
\usepackage[flushleft]{threeparttable}

\usepackage{algorithm}
\usepackage{algorithmic}
\usepackage{bm}
\usepackage{mathtools}
\usepackage{autonum}
\usepackage{tabularx}
\usepackage{enumitem}% http://ctan.org/pkg/enumitem
\usepackage{subfig}
\usepackage{soul}
\newcolumntype{R}{>{\raggedleft\arraybackslash}X}

\newcommand*{\MRRRATIO}{MRR_{ratio}}

\setcopyright{none}
\acmDOI{}

\acmISBN{}

\acmConference[]{}{}{}
\acmYear{}
\copyrightyear{}
\titlenote{This is an extended version of our work 'Intent term weighing in e-commerce queries', to appear in CIKM'19. The code for this paper is available at https://github.com/gurdaspuriya/query\_intent}

\begin{document}
\title{Intent term selection and refinement in e-commerce queries}
% \titlenote{Produces the permission block, and
\subtitle{}
% \subtitlenote{The full version of the author's guide is available as
%   \texttt{acmart.pdf} document}

\author{Saurav Manchanda}
\authornote{Work was done when Saurav was a summer intern at WalmartLabs.}
\affiliation{%
  \institution{University of Minnesota}
%   \streetaddress{,  55455, USA}
  \city{Twin Cities}
  \state{MN, USA}
  \postcode{55455}
}
\email{manch043@umn.edu}

\author{Mohit Sharma}
\authornote{Work was done when Mohit was at WalmartLabs.}
\affiliation{%
  \institution{WalmartLabs}
  \city{Sunnyvale}
  \state{CA}
  \postcode{94086}
}
\email{sharm163@umn.edu}

\author{George Karypis}
% \authornote{This author is the one who did all the really hard work.}
\affiliation{%
  \institution{University of Minnesota}
  \city{Twin Cities}
  \state{MN, USA}
  \postcode{55455}
 }
 \email{karypis@umn.edu}

\renewcommand{\shortauthors}{S. Manchanda et al.}

\begin{abstract}
In e-commerce, a user tends to search for the desired product by issuing a query to the search engine and examining the retrieved results.
If the search engine was successful in correctly understanding the user's query, it will return results that correspond to the products whose attributes match the terms in the query that are representative of the query's product intent.
However, the search engine may fail to retrieve results that satisfy the query's product intent and thus degrading user experience due to different issues in query processing: (i) when multiple terms are present in a query it may fail to determine the relevant terms that are representative of the query's product intent, and (ii) it may suffer from vocabulary gap between the terms in the query and the product's description, i.e., terms used in the query are semantically similar but different from the terms in the product description.
Hence, identifying the terms that describe the query's product intent and predicting additional terms that describe the query's product intent better than the existing query terms to the search engine is an essential task in e-commerce search.
In this paper, we leverage the historical query reformulation logs of a major e-commerce retailer to develop distant-supervised approaches to solve both these problems.
Our approaches exploit the fact that the significance of a term is dependent upon the context (other terms in the neighborhood) in which it is used in order to learn the importance of the term towards the query's product intent. We show that identifying and emphasizing the terms that define the query's product intent leads to a $3\%$ improvement in ranking.
Moreover, for the tasks of identifying the important terms in a query and for predicting the additional terms that represent product intent, experiments illustrate that our approaches outperform the non-contextual baselines.

\end{abstract}




\maketitle

\section{Introduction}
Online shopping has become a popular activity, with a recent report from the US Department of Commerce\footnote{https://www.census.gov/programs-surveys/arts.html} stating that the online sales grew $15.1\%$ in 2016 and accounted for $8.1\%$ of total retail sales for the year. With such a growth rate, improving the user experience with respect to the product search is a valuable challenge for the e-commerce companies. 

Retrieving the products that are relevant to a search query in an e-commerce site is a fundamental problem as it is often the first step a customer performs during an e-commerce transaction. The search engine relies on the terms in the query to return a set of products whose attributes match the terms in the query. Different terms in the search query describe the different characteristics of the product intent (relevant products to the query) such as brand, product-type, and other product attributes. For example, for the search query ``motorola phone'', the product-type intent is \emph{phone} and brand intent is \emph{motorola}. However, a search engine in e-commerce that relies on terms in a query for retrieval suffers from two different issues. First, when a query has multiple terms, it will retrieve all the products whose attributes match the terms in the query, but this affects the search results as it may also match the terms in the query that do not describe the query's product intent, e.g., the query ``socks for running shoes'' may return ``running shoes'' as well in addition to ``socks''. Moreover, the search engine suffers from the vocabulary gap when a term in the query that describes the intended product is not identical to the terms used in the description of the relevant products in the catalog, e.g., the term ``outdoor'' in query ``outdoor paint'' corresponds to the term ``exterior'' used often in product description of the relevant products.

The state-of-the-art term frequency based retrieval approaches, e.g., BM25 and BM25F, use the appearance of query terms in the product descriptions to compute a score indicating the relevance of the product for the query. 
However, a term that frequently occurs in the catalog will have a low IDF weight thereby lowering its contribution in the final score for queries having that term. This low contribution of term affects the recall of products as these terms can be the most critical terms in the query, e.g., the term ``women'' in ``women shoes'' is important but due to its high frequency and consequently low IDF weight,  the search engine may show ``men shoes'' leading to a bad customer experience followed by loss in sales and revenue. For the vocabulary gap problem, the existing approaches~\cite{chien2005semantic, boldi2008query,cucerzan2005extracting,sadikov2010clustering} rely on the historical engagement of a query to suggest terms for the already seen query and hence these approaches would not work for the rare queries.

In this paper, we use the query reformulation data derived from the historical search logs of a major e-commerce retailer\footnote{walmart.com}  to develop distant-supervised approaches~\cite{mintz2009distant} to solve these problems.  Specifically, we leverage the fact that the terms in the reformulated query describe the query's product intent in a better manner than the terms in the original query. Additionally, our approaches also take into account the context of a term, i.e., the entities in the neighborhood of the term,  in the query to estimate the representation of the query's product intent with Recurrent Neural Networks (RNNs).  For example, the same term \emph{3-piece} has different significance for the queries like ``3-piece kids dinnerware'' and ``3-piece mens suit''. For the first query, the context is \emph{kids dinnerware} and for the second query, the context is \emph{mens suit}.

In order to identify the relevant terms in a query that represent its intent, we estimate a weight for each term in the query that indicates the importance of the term towards expressing the query's product intent.  To estimate these term weights, we present a model that uses an \emph{intent encoder} to learn the query's product intent representation using RNNs, a \emph{feature generator} to create features that capture the contextual information, and a \emph{weight calculator} to estimate the weights for each term. Finally, to bridge the vocabulary gap between a query and the products relevant to the query we develop a query refinement approach to suggest terms that represent the query's product intent but are not in the original query. Our query refinement model uses an \emph{intent encoder} followed by a \emph{multilabel classifier} to predict the terms that express the query's product intent.

The approaches presented in this paper can be generalized to all the queries. However, from the e-commerce perspective, we can use the historical engagement, e.g., clicks, add-to-cart, and purchases, for the queries that have been issued by the users in the past to retrieve the relevant items for these queries. 
Therefore, we have restricted our evaluation to the rare queries, i.e., \emph{cold-start} or \emph{tail} queries, and our analysis shows that these queries constitute a significant chunk of the overall search traffic of the e-commerce retailer. The impact of the developed approaches will be more for the tail queries as we do not have any prior historical information to retrieve all the relevant items for these queries. 
To evaluate our approach on the term-weight prediction problem, we performed two tasks. The first one evaluates the ranking of the search results after weighting or boosting higher the terms that represent query's product intent. On the mean reciprocal rank (MRR)~\cite{voorhees1999trec} metric, our term-weight prediction approach improves the ranking by $3\%$ over the BM25F~\cite{robertson1995okapi} approach.
Considering that the e-commerce retailer's catalog contains millions of products, this improvement is significant and will help by generating a better set of initial candidate products followed by the application of Learning-to-Rank methods~\cite{karmaker2017application} to produce the final ranking.
The second task predicts the terms that are important towards defining the query's product intent and hence, are retained in the reformulated query. With respect to predicting the important terms, our approach performs better than the non-contextual baseline, with the relative performance gain of $6.7\%$ over the non-contextual baseline.
For query refinement, we evaluated our approach on predicting all the terms in the reformulated query, given the current query. With respect to predicting the terms in the reformulated query, our approach beats the non-contextual baseline with the relative performance gain of $3.4\%$.

The remainder of the paper is organized as follows. Section \ref{literature} presents the related literature review. Section \ref{definitions} presents the definitions and notations used in the paper. The paper discusses the developed methods in Section \ref{proposed} followed by the details of our experimental methodology in Section \ref{experiments}. Section \ref{results} discusses the results.  Finally, Section \ref{conclusion} provides some concluding remarks.

\section{Related work}\label{literature}
In this section, we present a brief overview of the prior literature relevant to the work presented in this paper and discuss their limitations. We look at the term-weighting methods, methods that improve relevance for the tail queries and query refinement methods.
\subsection{Term-weighting methods}\label{section:term-weight}
Feature selection using term-weighting is a well-researched topic in information retrieval, but the prior work in this area has been regarding ranking terms according to their discriminating power in determining relevance, the relevance based on the text-matching between the query and document (or product). Popular examples of such term-weighting methods include term frequency-inverse document frequency (TF-IDF)~\cite{salton1986introduction} and Okapi Best Matching 25 (Okapi BM25)~\cite{robertson1995okapi, robertson1994some}. Intuitively, such techniques determine the importance of a given term in a particular document. Terms that are common in a single or a small group of documents tend to have higher weights than frequent terms. Such term-weighting techniques are well-suited to rank documents in response to a query, but cannot be applied on a group of queries to rank the terms in individual queries with respect to their confidence in defining the query's product intent. A term might be discriminatory with respect to a query (that is, the term occurs in a small number of queries), but this does not necessarily mean that the term is important towards defining that query's product intent. For example, in the query ``striped mens shirt'', the terms \emph{mens} and \emph{shirt} are expected to be more frequent than the term \emph{striped}. The above weighting techniques will tend to give higher weight to \emph{striped}, but we expect that the terms \emph{mens} and \emph{shirt} should carry more weight towards defining the query's product intent.

Another relevant approach that can be used to estimate term-weights is to first construct a click graph between the queries and documents and estimate a vector representation for each entity (queries and documents) using a vector propagation model (VPCG \& VG)~\cite{jiang2016learning}. The queries or documents can then be broken down into individual units (e.g., $n$-grams), and we can learn a vector representation for each $n$-gram based on the vectors already estimated from the click graph. We can then estimate a weight for each $n$-gram by a regression model that minimizes the square of the euclidean distance between the representation of a query vector and the weighted linear combination of the representation of the $n$-grams in the query. The limitation of this approach is that it estimates a global weight for each $n$-gram, and hence does not take care of the fact that the significance of a term ($n$-gram) is dependent upon the context in which it is used. 

\subsection{Improving relevance for the tail queries}
Another class of methods that are relevant to the work present in this paper is the work regarding understanding the tail queries~\cite{song2014transfer, downey2007heads}. Queries can be classified into head queries (frequent queries, associated with rich historical user engagement information) and tail queries (rare queries, do not have much historical user engagement data). Head queries enable retrieval engines to utilize statistical models to correctly identify the query's product intent from the historical engagement data (clicks, add to carts, and  orders). Tail queries constitute the majority of unique queries submitted by the users, and hence, it is very valuable challenge for e-commerce companies to improve search relevance for tail queries. 

One of the proposed approaches to improve the relevance for the tail queries is to predict the likelihood of one query being the reformulation of another query~\cite{downey2007heads}. The above approach is based on Probabilistic Latent Semantic Analysis
(PLSA). It models each pair of consecutive queries that a user composes as generated by a single latent topic. As such, the above proposed model makes sense when we are looking to find the most likely (or a set of most likely) reformulation given a query and the set of candidate reformulations is small. But the reformulations cannot be limited to a small set of head queries.

Another approach to improve search relevance for tail queries is to extract single or multi-word expressions from both the head and tail queries and mapping them to a common concept space defined by a knowledge-base~\cite{song2014transfer}. Once both the head and tail queries are mapped to the same concept space, historical engagement data for the head queries can be used for the tail queries as well via the mapped concepts. The limitation of this approach is that mapping tail queries to a logical concept space using query text is again prone to noisy terms, and the concepts estimated from the noisy terms will degrade the performance. 

The VPCG \& VG~\cite{jiang2016learning} approach described in Section \ref{section:term-weight} can also be used to improve the relevance of tail queries. From the estimated weights and vectors of the $n$-grams, we can estimate the vectors for the tail queries by a linear combination of the vectors of the constituent $n$-grams. As mentioned before, the limitation of this approach is that it does not take care of the fact that the significance of a term is dependent upon the context in which it is used. 

\subsection{Query refinement}
Query refinement (QR) refers to reformulating a given query to improve its search relevance. Query refinement approaches can be either knowledge-based (using knowledge-bases to find semantically related terms) or data-based (statistical models using historical data, like search logs and engagement information). Examples of query refinement include query expansion and query suggestion~\cite{zhang2016towards}. The query refinement work presented in this paper is data-based and uses historical search logs to build a statistical model.

One of the most popular approaches for query refinement is the use of relevance feedback~\cite{rocchio1971relevance} where the term suggestions (suggesting terms relevant to the query's product intent) are based on the user engagement with the results retrieved for the initial query. In the absence of user feedback, term suggestion can be done via pseudo relevance feedback, which assumes that the top retrieved results are relevant to the query's product intent. These relevance based approaches assume that the retrieval system is able to understand the query's product intent and retrieve relevant results to some extent, which is an invalid assumption for tail queries with noisy terms.

A number of prior approaches have been proposed to use historical search logs for query refinement. The prior approaches work by finding similarity between the queries, so that, given a query, other similar queries can be used to refine it. The reformulateed queries from historical logs can be clustered to identify different topics or aspects of a query~\cite{radlinski2010inferring,dang2011inferring}. \citet{chien2005semantic} use the temporal correlation in sessions between two queries as the similarity between queries. \citet{cucerzan2005extracting, boldi2008query} use the session co-occurrence information between two queries to find the similarity between them. \citet{sadikov2010clustering} also use the document-click information in addition to co-occurrence information between two queries for query recommendation. A common drawback of the above approaches is that they cannot work on unseen queries, and hence are not helpful with tail queries, which are mostly unique and issued rarely.

One of the approaches analyzes the relations of terms inside a query and
uses the discovered term association patterns for effective query reformulation~\cite{wang2008mining}. Since this approach is based on the co-occurrence patterns of the terms, it can be applied on the tail queries. It estimates a context distribution for terms occurring in a query and then suggests similar words based on their distributional similarity. The drawback of this approach is that the term association patterns are discovered from the co-occurrence statistics of the terms within a query, independent of actual reformulations of that query. Therefore, the discovered patterns lack the task-specific context of reformulation.
\section{Definitions and Notations} \label{definitions}

\begin{table}[!t]
\small
\centering
  \caption{Notation used throughout the paper.}
  \begin{tabularx}{\columnwidth}{lX}
    \hline
Symbol   & Description \\ \hline
$V$    & Vocabulary (set of all the terms). \\
$q$    & Query such that the results displayed do not satisfy the query's intent. \\
$|q|$    & Number of terms in $q$ \\
$R(q)$    & Reformulation of $q$ such that the results displayed satisfy the query's intent. \\
$Q$    & Collection of all the queries $q$ \\
$\mathbf{r}_x$    & Word vector representation of the word $x$ \\
$d$    & Length of the word vector representation \\
$\mathbf{h}^f_t$    & Hidden state at position $t$ for the $GRU^f$ \\
$\mathbf{h}^b_t$    & Hidden state at position $t$ for the $GRU^b$ \\
$l$    & Size of the hidden states for the GRUs \\
$s(x)$    & Importance weight for the term $x$ \\
\hline
\end{tabularx}
  \label{tab:notation}
\end{table}
Let $V$ be the set of distinct terms that appear in all the queries of a website, referred to as the \emph{vocabulary}. Let $q$ be a query such that the results (products) displayed for $q$ do not satisfy its product intent. The query $q$ is a sequence of $|q|$ terms from the vocabulary $V$, that is,
\begin{equation}
q = \langle v_{q_{1}}, \ldots, v_{q_{|q|}}\rangle.
\end{equation}
The set of all such queries $q$ is denoted by $Q$.  Let $R(q)$ denote a reformulation of $q$ such that the results displayed for $R(q)$ satisfy its product intent. 
Each word $u_i$ in the vocabulary will also be represented by a $d$-dimensional vector $r_{u_i}$.

\begin{figure*}
\includegraphics[width=\textwidth]{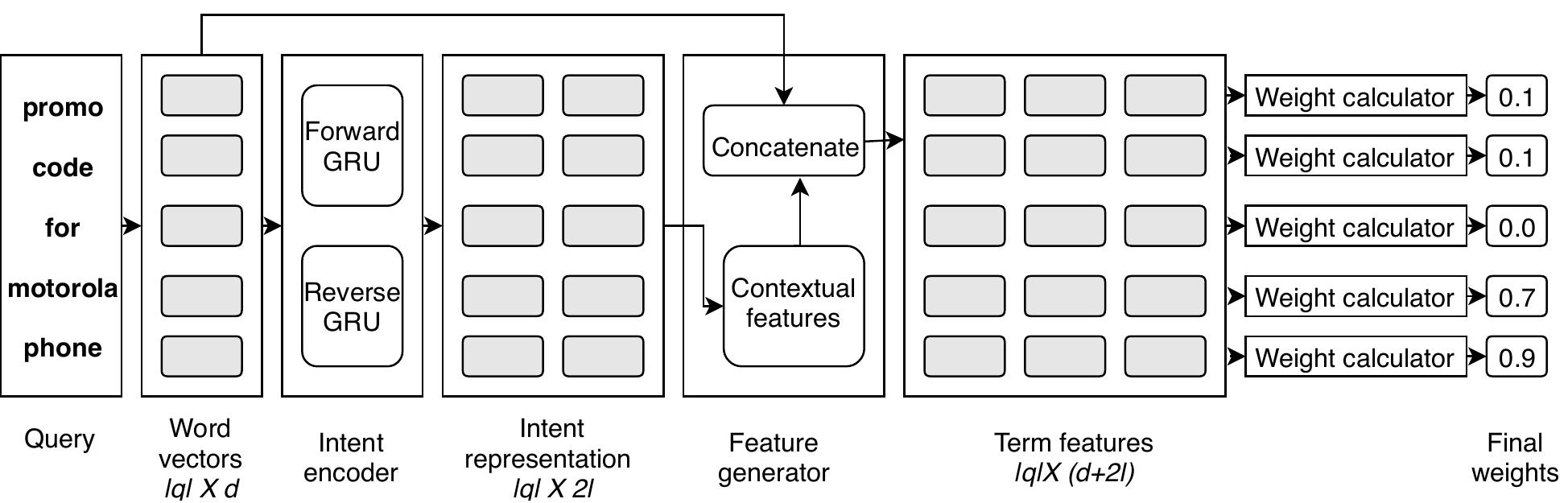}
\caption{Contextual term-weighting (CTW) model.}
\label{fig:term_weighting}
\end{figure*}

Table \ref{tab:notation} provides a reference for the notation used throughout the paper.

\section{Proposed methods}\label{proposed}

In order to identify the terms that express a query's product intent, we do not rely on access to a labeled data but use the query reformulation logs to develop a distant-supervised approach to find terms that represent the query's product intent.  
Specifically, we assume that we are given a query $q$ and its reformulation $R(q)$ such that the results displayed for $q$ does not satisfy its product intent while results displayed for $R(q)$ satisfy the product intent. We assume that the terms in $q$ that represent its product intent are also retained in $R(q)$, i.e., the terms in set $q \cap R(q)$ are critical in expressing $q$'s product intent than the terms in set $q \setminus R(q)$. 
Our approach \emph{contextual term-weighting (CTW)} model solves this problem by predicting a weight for each term in a query that indicates the importance of that term towards defining the query's product intent. CTW leverages the context of a term, i.e., the entities in the neighborhood of the term, to estimate its weight. CTW comprises of an \emph{intent encoder} that estimates the representation of a term in query with Recurrent Neural Networks (RNNs), a \emph{feature generator} that uses the intent representation to generate features that capture the contextual information, and a \emph{weight calculator} that uses multilayer perceptron (MLP) to estimate the weights for each term using the generated features. 

Additionally, we focus on addressing the vocabulary gap between the terms in a query and a relevant product's description, i.e., terms used in the query are semantically similar but different from the terms in the description of the product. For example,  for query  ``outdoor paint'' a relevant product's description contains semantically similar term instead of ``outdoor'' in its title, i.e., ``exterior paint''. 
We want to refine a query by suggesting other terms that are not present in the query but express its product intent better than the original terms in the query. We assume that the terms in $R(q)$ are candidates for the query refinement of $q$ as the terms in the reformulated query define the product intent in a better way than query $q$. Our approach \emph{Contextual query refinement (CQR)} model suggests relevant terms by classifying each term in the vocabulary if the term defines the $q$'s product intent. Similar to CTW, CQR also leverages the context of a term to make its predictions. CTW comprises of an \emph{intent encoder} that estimates the product intent of the query with Recurrent Neural Networks (RNNs) and a \emph{multilabel classifier} that classifies each term if it is relevant towards defining the query's product intent.

\subsection{Contextual term-weighting (CTW) model}
Given a query $\langle v_{q_{1}}, \ldots, v_{q_{|q|}}\rangle$, our task is to estimate a weight $s(v_{q_{t}})$ for each term $v_{q_{t}}$, which captures how important is $v_{q_{t}}$ towards defining $q$'s product intent. The supervision during training comes from the fact that the terms in the set $q \cap R(q)$ are more important in defining the product intent than the terms in the set $q \setminus R(q)$.

Figure \ref{fig:term_weighting} provides an illustration of the contextual term-weighting model. The contextual term-weighting model has three modules: \emph{intent encoder}, \emph{feature generator} and \emph{weight calculator}. The \emph{intent encoder} takes $q$ as input, and outputs a representation of the product intent of $q$. Next, the \emph{feature generator} uses the output of the \emph{intent encoder} as input and generates the features for each query term $v_{q_{t}}$, which captures its importance towards the product intent of $q$. For each term $v_{q_{t}}$, the \emph{weight calculator} uses the features produced by the \emph{feature generator} as input and outputs the importance weight for it. We discuss these three modules in detail below:

\begin{figure*}
\includegraphics{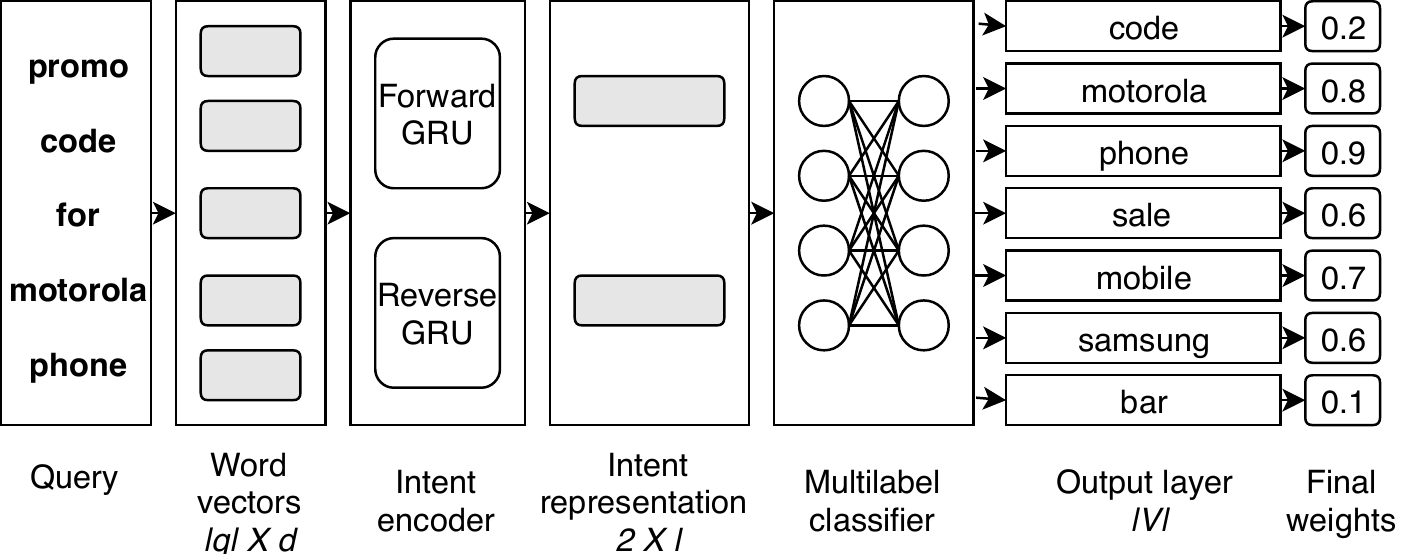}
\caption{Contextual query refinement (CQR) model.}
\label{fig:query_refinement}
\end{figure*}
\par \emph{\textbf{Intent encoder}}: This module encodes the sequence of terms in the query $q$ into a fixed length representation using bidirectional Gated Recurrent Units (GRUs)~\cite{cho2014properties}. GRUs are a type of Recurrent Neural Networks (RNNs), which model variable-length sequential input using a recurrent, shared hidden state. The hidden state can be thought of a summary of the complete sequential input. The sequence of word vectors in the query $q$ is denoted by $\langle \mathbf{r}_{v_{q_{1}}, \ldots, \mathbf{r}_{v_{q_{|q|}}}} \rangle$. The bidirectional GRU encodes the query $q$ as: 
\begin{equation}\label{eq:gruf}
\mathbf{h}^f_t = GRU^f(\mathbf{h}^f_{t-1}, \mathbf{r}_{v_{q_{t}}})
\end{equation}
\begin{equation}\label{eq:grub}
\mathbf{h}^b_t = GRU^b(\mathbf{h}^b_{t+1}, \mathbf{r}_{v_{q_{t}}}),
\end{equation}
where $GRU^f$ encodes the query $q$ in the forward direction and $GRU^b$ encodes in the backward direction. $\mathbf{h}^f_t$ is the hidden state at position $t$ for the $GRU^f$ and corresponds to the summary of the sequence $\langle \mathbf{r}_{v_{q_{1}}}, \ldots, \mathbf{r}_{v_{q_{t}}}\rangle$. Similarly, $\mathbf{h}^b_t$ is the hidden state at position $t$ for the $GRU^b$ and stores to the summary of the sequence $\langle \mathbf{r}_{v_{q_{t}}}, \ldots, \mathbf{r}_{v_{q_{|q|}}}\rangle$. The output of the intent encoder are the hidden states ($\mathbf{h}^f_t$ and $\mathbf{h}^b_t$) at each position $t$. 

\par \emph{\textbf{Feature generator}}: This module generates features for each term $v_{q_{t}}$ in the query $q$, the features capturing its importance towards defining the $q$'s product intent. The importance of a term is not just dependent on the term itself, but also on the context in which the term is used. For example, the same term \emph{3-piece} can have different importance for the different queries like ``3-piece kids dinnerware'' and ``3-piece mens suit''. 

The term level features can be captured using the word vectors. To capture the contextual features of a term, we use the output from the \emph{intent encoder} module. Consider the forward encoder $GRU^f$. At each position $t$, $GRU^f$ updates the current summary $\mathbf{h}^f_{t-1}$ with the term $v_{q_{t}}$ at point $t$. The contribution of the term $v_{q_{t}}$ towards defining the product intent of the query should be manifested in the extent to which it updates the summary at point $t-1$. Therefore, the contribution of a term at position $t$ towards defining the product intent of the query should be a function of the difference in the summaries at the positions $t$ and $t-1$ ($\mathbf{h}^f_{t} - \mathbf{h}^f_{t-1}$). Similarly, for the reverse encoder $GRU^b$, it should be a function of the difference in the summaries at the positions $t$ and $t+1$ ($\mathbf{h}^b_{t} - \mathbf{h}^b_{t+1}$).

For each term $q_t$ at position $t$ in the query, the output of the feature generator ($f_t$) is the concatenation of the vectors $\mathbf{r}_{v_{q_{t}}}$, $\mathbf{h}^f_{t} - \mathbf{h}^f_{t-1}$ and $\mathbf{h}^b_{t} - \mathbf{h}^b_{t+1}$.

\begin{equation}
f_t = [\mathbf{r}_{v_{q_{t}}}, \mathbf{h}^f_{t} - \mathbf{h}^f_{t-1}, \mathbf{h}^b_{t} - \mathbf{h}^b_{t+1}].
\end{equation}

\par \emph{\textbf{Weight calculator}}: This module takes as input the features $f_t$ generated by the feature generator for a term $v_{q_{t}}$, and outputs a weight for that term, the weight manifesting the importance the term $v_{q_{t}}$ has towards defining the product intent of the query $q$.

We model the weight calculator as a multilayer perceptron (MLP) with a single node at the output layer. The \emph{weight calculator} outputs a weight between $0$ and $1$. We refer to the weight corresponding to the query term $v_{q_{t}}$ as $s(v_{q_{t}})$.

To train the model, we minimize the binary cross entropy loss~\cite{nam2014large}. Using the binary cross entropy loss allows our model to make predictions independently for each term, that is, predicting high weight for one term will not effect the weights of the other terms. The binary cross entropy loss associated with all the queries in the collection $Q$ is given by:
\begin{equation}
L(Q) = -\sum_{q\in Q}\sum_{t=1}^{|q|} (y(v_{q_{t}}) \log(s(v_{q_{t}})) + (1 - y(v_{q_{t}}))\log(1 - s(v_{q_{t}}))),
\end{equation}
where $y(v_{q_{t}}) = 1$ if the term $v_{q_{t}}$ is also present in the reformulated query $R(q)$, and $y(v_{q_{t}}) = 0$ otherwise.

\subsection{Contextual query refinement (CQR) model}

Given a query $\langle v_{q_{1}}, \ldots, v_{q_{|q|}}\rangle$, our task is to find and recommend terms $v_i$ from the vocabulary $V$, which define the product intent of the query $q$, and hence can be used to refine $q$. The supervision during training comes from the fact that the terms in the reformulation $R(q)$ define the product intent better than the terms in the original query $q$, as the products displayed for $R(q)$ satisfied the product intent. We model the query refinement as a multilabel classification model, the label space being the set of all the terms in the vocabulary $V$.

Figure \ref{fig:query_refinement} provides an illustration of the contextual query refinement model. The query refinement model has two modules: \emph{intent encoder} and \emph{multilabel classifier}. The \emph{intent encoder} takes as input the query $q$, and outputs a representation of the product intent of the query $q$. Next, the \emph{multilabel classifier} uses the output of \emph{intent encoder} as input and generates a weight $s(v_i)$ for each term $v_i$ in the vocabulary $V$, the weight establishing the confidence of that term towards defining $q$'s product intent. We discuss these two modules in detail below:

\par \emph{\textbf{Intent encoder}}: Similarly to the query term-weighting model, we encode query $q$ using a bidirectional GRU using the equations \ref{eq:gruf} and \ref{eq:grub}.

However, unlike the intent encoder for weighting query terms, we are only interested in the product intent representation of the complete query, not the representation at individual positions in the query. For the forward encoder $GRU^f$, the hidden state at the position $|q|$ captures the product intent of the query $q$. Similarly, for the reverse encoder $GRU^b$, the hidden state at the position $1$ captures the intent of the query $q$. We use the concatenation of $\mathbf{h}^f_{|q|}$ and $\mathbf{h}^f_{1}$ as the overall intent representation of the query $q$, and represent it by $\mathbf{h}_q$
\begin{equation}
\mathbf{h}_q = [\mathbf{h}^f_{|q|}, \mathbf{h}^b_{1}].
\end{equation}

\par \emph{\textbf{Multilabel classifier}}: This module takes the the product intent representation of the query $q$ as input and outputs a weight for each term $v_i$ in the vocabulary $V$, the weight manifesting the importance the term $v_i$ has towards defining the product intent of the query $q$.

We model the multilabel classifier as a multilayer perceptron (MLP) with output layer having the number of nodes equal to the size of vocabulary ($|V|$). For every term $v_i$ in the vocabulary, the multilabel classifier outputs a weight between 0 and 1. We refer to the weight corresponding to the term $v_i$ as $s(v_i)$.

To estimate the model, we minimize the binary cross entropy loss. The binary cross entropy loss associated with all the queries in the collection $Q$ is given by:
\begin{equation}
L(Q) = -\sum_{q\in Q}\sum_{i=1}^{|V|} (y(v_i) \log(s(v_i)) + (1 - y(v_i))\log(1 - s(v_i))),
\end{equation}
where $y(v_i) = 1$ if the term $v_i$ is present in the reformulated query $R(q)$, and $y(v_i) = 0$ otherwise.

Both the models, i.e., Contextual term-weighting (CTW) model and Contextual query refinement (CQR) model, can generalize to unseen tail queries that have the same vocabulary as that of the queries in historical search logs.

\section{Experimental methodology}\label{experiments}
\subsection{Dataset}

To understand the query reformulation patterns in product search, we analyzed the historical search logs of a major e-commerce retailer\footnote{walmart.com} and looked at the query transitions of the form $a \rightarrow b$, where $a$ and $b$ are two queries within the same session and $b$ is issued immediately after issuing $a$. The transitions $a \rightarrow b$ broadly fall into the following five categories:
\begin{description}[style=unboxed,leftmargin=0cm]
    \item[Transition from a general to specific product intent:] For example, $a$ is \emph{furniture} and $b$ is \emph{wooden furniture}. The query $a$ in this category is usually short and the set of the terms in $a$ is usually a proper subset of the set of terms in $b$.
    
    \item[Transition from an incomplete to a complete query (user pressed enter key before finishing the query):] For example, $a$ is \emph{air condi} and $b$ is \emph{air conditioner}. The query $a$ in this category tends to have spelling errors.
    
    \item[Change of intent:] For example, $a$ is \emph{bedside lamps} and $b$ is \emph{light bulbs}. The terms in $a$ tend to be different than the terms in $b$.
    
    \item[Transition from a specific to general product intent:] For example, $a$ is \emph{3-piece kids dinnerware} and $b$ is \emph{kids dinnerware}. 
    
    \item[Reformulation with the same product intent:] For example, $a$ is \emph{promo code for motorola phone} and $b$ is \emph{motorola phone on sale}. 
\end{description}

We need to extract query pairs of the form $(q, R(q))$ such that the user is unable to find the intended product from the results retrieved for the query $q$. Within the same session, he reformulates the query to $R(q)$, and is able to find the required products from the results displayed for $R(q)$. 

We assume that the user's engagement with the displayed products (click, add-to-cart, order) is a proxy for the satisfaction of the query's product intent. However, lack of user's engagement does not necessary mean that the displayed products do not satisfy the query's product intent (for example, $a$ is \emph{furniture} and $b$ is \emph{wooden furniture}). However, for the following two types of transitions, we can say that the product intent was not satisfied for the initial query: \emph{Transition from a specific to general product intent} and \emph{Reformulation with the same product intent}.  We can manually look into the query reformulations to collect the required query pairs, but this approach is expensive in terms of both time and money. Instead, we generate the required query pairs by performing the following steps:
\begin{itemize}
    
    \item We sampled an initial dataset of query pairs $(a, b)$ from the historical search logs data spanning over eleven months (July'17 to May'18), such that user searches for $b$ after query $a$, and the user searches at most two other queries between $a$ and $b$. 
    
    \item We only keep the pairs for which there is an add to cart (ATC) event associated with $b$. This restricts $b$ to the queries for which the product intent is satisfied.
    
    \item We limit $a$ to only rare queries (less than 300 occurrences over a 60-day period and the fraction of times user clicks (click through rate) any product displayed for $a$ is less than 5\%). Note that, we limit ourselves to the rare queries because the frequent queries with small click-through-rate can be attributed to crawling by bots, which is a common practice in e-commerce by online retailers to monitor prices, product descriptions etc.
    
    \item We only look into the pairs such that the Jaccard similarity~\cite{jaccard1901etude} between the sets of terms in $a$ and $b$ is at least $0.2$. This ensures that the product intent does not change drastically between the queries. 
    
    \item We limit our dataset to contain only those $a$'s whose constituent terms' frequency is more than 100 over the complete dataset (if some term is frequent, it is probably not a spelling error). This filters out the query pairs for which $a$'s product intent is not satisfied as a result of spelling errors.
    
    \item Since the lack of user's engagement does not necessary mean that the displayed products do not satisfy the query's product intent (when the initial query has a very general product intent, for example, the initial query is \emph{furniture} and the reformulated query is \emph{wooden furniture}), we limit our dataset to the pairs, such that the set of terms in $a$ is not a proper subset of the set of terms in $b$. Also we only keep pairs with $a$ having at least three terms, as the queries with a general product intent tend to be shorter. 
\end{itemize}

From the remaining pairs, we create a training-test-validation split. We have 722,235 query pairs in the training set, 50,000 query pairs in the test set and 5,000 pairs in the validation set. The vocabulary contains 12,118 terms.

Note that the approaches developed by us can be generalized to all the queries. However, from the e-commerce perspective, we can use the historical engagement, e.g., clicks, add-to-cart, and conversions, for the queries that have been issued by the users in the past to retrieve the relevant items for these queries. For the same reason, we constructed our dataset to only contain the reformulations where the initial query is a rare query. Hence, we have restricted our training and evaluation to the queries that are not frequent, i.e., \emph{cold-start} or \emph{tail} queries, and these queries constitute a significant chunk of the overall search traffic of the e-commerce retailer. The impact of the developed approaches will be more for the tail queries as we do not have any prior historical information to retrieve all the relevant items for these queries. 
\subsection{Evaluation Methodology and Performance Assessment}
\subsubsection{Methodology and metrics}
To evaluate our approach on the term-weight prediction problem, we performed two tasks.
We describe our evaluation methodology and metrics for the term-weighting and query refinement problems below.
\par \emph{Term-weighting:} We need to evaluate how well the intent term weighting improves the ranking of the search results in response to a query. To apply the intent term weighting for ranking products in response to a query, we first retrieve the top products for the query using a BM25F~\cite{robertson1995okapi} based retrieval algorithm. The original relevance score of each product (without the intent term weighting) is calculated as the sum of the corresponding BM25F scores for each term in the query. Then, the individual BM25F score is scaled (boosted) with the corresponding computed term-weight and re-ranking is performed with the modified scores. We calculate the Mean Reciprocal Rank (MRR), which is the average of the reciprocal of the rank of the relevant product in response to a query. 
The relative performance improvement in MRR is given by
\begin{equation}
    \MRRRATIO = \frac{MRR_{boost}}{MRR_{BM25F}},
\end{equation}
\noindent where $MRR_{boost}$ is the MRR when we have scaled the BM25F scores of each term with the corresponding term weight, and $MRR_{BM25F}$ is the MRR when we have used the BM25F scores of each term as it is.

Moreover, we also evaluated our approach on how well it is able to estimate weights for the query terms in order of their importance towards defining the product intent of the query. We assume that, if a term is important, it should be present in the reformulated query too, for which the displayed products satisfy the product intent. Therefore, to evaluate our approach on the term-weighting problem, we looked at how well our model is able to estimate higher weights for the terms in the set $q \cap R(q)$ than the terms in the set $q \setminus R(q)$.

For this prediction task, we used Precision@$k$ ($P@k$), which is a popular metric used in multilabel classification literature~\cite{agrawal2013multi}. Given a list of ground truth labels, ranked according to their relevance, $P@k$ measures the precision of predicting the first $k$ labels from this list. Since we cannot score ground truth labels for the term-weighting task, that is, the ground truth labels are binary, we use $k = nnz$, where $nnz$ is the output sparsity of the ground truth labels. We also ignored the stop words in the calculation of $P@k$. For example, lets say the initial query is ``promo code for motorola phone'' and the reformulated query is ``motorola phone on sale''. In this case, the words \emph{motorola} and \emph{phone} are retained in the reformulation, and hence we use $nnz=2$ for the term-weighting problem. We report the averaged $P@nnz$ ($AP@nnz$) over all the test instances. We also present results for $k = 1, 2, 3$ as the queries in product search tend to be shorter (the average length of the reformulated query in our test set is $3.6$).
\par \emph{Query refinement:} The terms that appear in the reformulation $R(q)$, for which the query's product intent is satisfied, describe the product intent in a better manner than the terms in the initial query $q$. Therefore, if we can predict the terms in the reformulation, we can refine the query with these predicted terms. We evaluated our query refinement approach on how well it is able to predict the terms in the reformulated query for which the displayed products satisfy the product intent. Similar to the term-weighting task, we report the averaged $P@nnz$ ($AP@nnz$) over all the test instances. $P@nnz$ in this case measures how many of the top $nnz$ predicted terms also appear in the reformulated query. Lets say the initial query is ``promo code for motorola phone'' and the reformulated query is ``motorola phone on sale''. For the query refinement problem, we use $nnz=3$ corresponding to three non-stop words in the reformulated query (\emph{motorola}, \emph{phone} and \emph{sale}). We also present results for $k = 1, 2, 3$. 

\subsubsection{Baselines}
Our approaches use the contextual information for term-weighting and query refinement. To illustrate the advantage of using contextual information, we compared our approaches against the baselines that do not take the context into consideration.

\par \emph{Term-weighting: } Given a query $q$ and a term $v_{q_t}$ in $q$, $s(v_{q_t})$ denotes the confidence of word $v_{q_t}$ towards defining the $q$'s product intent. As mentioned earlier, we assume that if the term $v_{q_t}$ has high confidence towards defining the product intent of $q$, $v_{q_t}$ should also be present in the reformulated query $R(q)$. Hence, we can calculate $s(v_{q_t})$ from the historical data as follows:
\begin{equation}
s(v_{q_t}) = P(v_{q_t}\in R(q) | v_{q_t} \in q)
\end{equation}

\begin{equation}
s(v_{q_t}) = \frac{P(v_{q_t}\in R(q), v_{q_t} \in q)}{P(v_{q_t} \in q)}.
\end{equation}

Using the maximum likelihood estimation, we have,

\begin{equation}
s(v_{q_t}) = \frac{\sum_{q'\in Q}1(v_{q_t}\in R(q'), v_{q_t} \in q')}{\sum_{q'\in Q}1(v_{q_t} \in q')},
\end{equation}

where $\sum_{q'\in Q}1(v_{q_t}\in R(q'), v_{q_t} \in q')$ denotes the number of times the term $v_{q_t}$ occurs in any query $q' \in Q$ and is retained in the reformulated query $R(q')$. $\sum_{q'\in Q}1(v_{q_t} \in q')$ denotes the number of times $v_{q_t}$ occurs in the initial query $q'\in Q$. We call this method of estimating query term-weights frequentist term-weighting (FTW).

As discussed in Section \ref{literature}, term-weighting methods that rank terms according to their discriminating power are not well suited to rank terms according to their contribution towards the product intent of the query. A term might be discriminatory with respect to a query (that is, the term occurs in a small number of queries), but this does not necessarily mean that the term is important towards defining that query’s product intent.  To illustrate this, we also compared our approach against term frequency-inverse document frequency (TF-IDF). 

As a competing approach, we also compared our approach against one of the methods that uses use click-through logs over query-product pairs to estimate vector representations for both query and product in the vocabulary space of search queries. We use the vector propagation method (VPCG) to learn the vector representation of search queries and the products in the same space~\cite{jiang2016learning}. Next, we learn the representation of the $n$-grams (uni- and bi-grams) present in the queries by a weighted sum of the representation of the queries having these $n$-grams.  We estimate the vector representation of a new query by a weighted linear combination of the representation of the n-grams present in the queries  (VPCG \& VG).  
These weights of the individual $n$-grams are estimated by minimizing the square of the euclidean distance between the representation of a query vector and the weighted linear combination of the representation of the $n$-grams in the query.
The relevance score of a product for a query is calculated by the scalar product of the vector representations of the product and the query. Note that we can use the estimated weights for each unigram as the importance weight for that term, but there is only one weight estimated for each term, and hence, this approach does not take the context into consideration.

\par \emph{Query refinement:} For the query refinement problem, we need to calculate the confidence of each term in the vocabulary towards defining the product intent of the query. Given any word $v_i$ in the vocabulary, let $s(v_i)$ denote the confidence of the word $v_i$ towards defining the product intent of the query $q$. We define $s(v_i)$ as the sum of probabilities of the term $v_i$ appearing after each of the terms $v_{q_t}$ in the query $q$, that is, we calculate $s(v_i)$ as,

\begin{equation}
s(v_i) = \sum_{v_{q_t} \in q}P(v_i\in R(q) | v_{q_t} \in q)
\end{equation}

\begin{equation}
s(v_i) = \sum_{v_{q_t} \in q}\frac{P(v_i\in R(q))}{P(v_{q_t}\in q)} 
\end{equation}

\begin{equation}
s(v_i) = \sum_{v_{q_t} \in q}\frac{\sum_{q'\in Q}1(v_i\in R(q'), v_{q_t} \in q')}{\sum_{q'\in Q}1(v_{q_t} \in q')}
\end{equation}

We call this method of finding query refinement terms frequentist query refinement (FQR). FQR can be thought of discovering term association patterns as proposed by~\citet{wang2008mining}. However, instead of discovering these patterns from the co-occurrence statistics of the terms within a query, we discover them from the query reformulations, thus making these patterns task-specific for a fair evaluation.

\subsubsection{Parameter selection}
We explored different hyper-parameters on validation set to select our hyper-parameters. For both term-weighting and query refinement, we used 300-dimensional word vectors as input to the GRUs in the intent encoder. We initialized the word-vectors using pre-trained skip-gram vectors~\cite{mikolov2013distributed} trained on one year of query data. We used 2-layer GRUs for both forward and reverse encoding, with 256-dimensional hidden layer. For regularization, we used a dropout~\cite{srivastava2014dropout} of $0.25$ between the first and second layers of both the GRUs. For the \emph{weight calculator} module in term-weighting, we used a two-layer multilayer perceptron (MLP) with 10 nodes in the hidden layer. For the \emph{multilabel classifier} module in query refinement, we used a two-layer multilayer perceptron (MLP) with $2\times |V|$ nodes in the hidden layer (there are $|V|$ nodes in the output layer). For both these MLPs, we used a dropout of $0.25$ between the input and hidden layer. We applied Rectified Linear Unit (ReLU) non-linearity on the hidden layer and sigmoid function on the output layer nodes.

For both the models, we used ADAM~\cite{kingma2014adam} optimizer for optimization with $0.001$ as the initial learning rate and a batch-size of $512$. We trained both the models for $20$ epochs. 

For VPCG \& VG, we experimented with different dimensions of the vector representations and selected the final value for the dimension of vector representations i.e., $50$, based on the ranking performance on a validation set. We ran the vector propagation method for maximum iterations, i.e., $50$, or until the ranking performance does not improve on the validation set.  We used stochastic gradient descent to learn weights for different $n$-grams and used $0.001$ as the value of our learning rate.

\section{Results and Discussion}\label{results}
\subsection{Quantitative evaluation}

\begin{table}[!t]
\small
\centering
  \caption{Results for the query term-weighting problem.}
  \begin{threeparttable}
  \begin{tabularx}{\columnwidth}{Xrrrrr}
    \hline
    Model    & $\MRRRATIO$      & $AP@nnz$  & $AP@1$  & $AP@2$  &$AP@3$\\ \hline
    CTW{*} & \textbf{1.030} & \textbf{0.746} & \textbf{0.792} & \textbf{0.804} & \textbf{0.832}       \\
    FTW & 1.019 & 0.699 & 0.746 & 0.750 & 0.790       \\
    TF-IDF   & 1.006 & 0.548 & 0.514 & 0.603 &0.700      \\
    VPCG \& VG   & 0.694 & 0.558 & 0.537 & 0.616 &0.709      \\ \hline
  \end{tabularx}
  \begin{tablenotes}
     \item[*] The performance of the approach was found to be statistically significant ($p$-value $< 0.05$ using paired $t$-test).
  \end{tablenotes}
  \end{threeparttable}
  \label{tab:term_weighting}
\end{table}

\subsubsection{Term-weighting} 
\begin{figure}
\includegraphics[width=\linewidth]{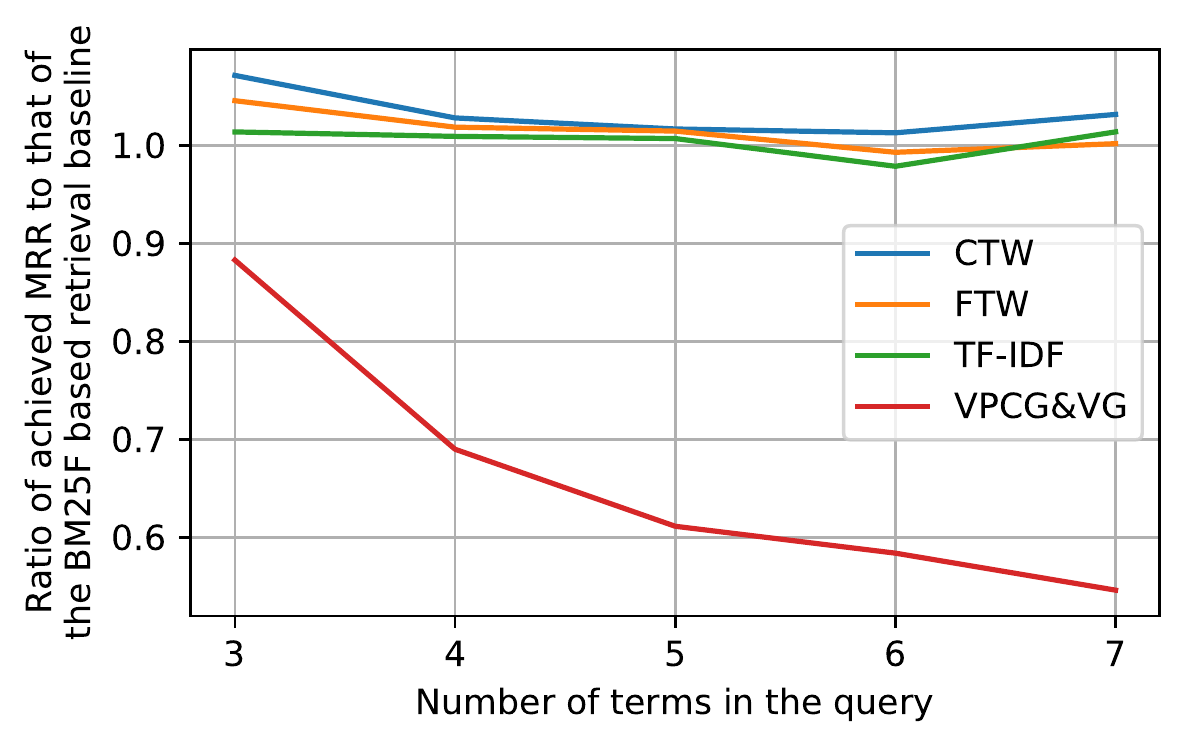}
\caption{Variation of the MRR with query length.}
\label{fig:mrr_length}
\end{figure}

Table~\ref{tab:term_weighting} shows the performance statistics for the term-weighting problem for the various methods. The intent term weighting performed by our proposed method CTW, FTW and TF-IDF leads to better ranking of the search results, as shown by the improvement on the MRR metric. CTW leads to the maximum increase in the MRR, achieving an improvement of $3\%$ against the case when no intent term weighting is applied. This illustrates the advantage of using contextual knowledge to find important terms from a search query. We found the difference between the reciprocal ranks (RR) for CTW and FTW to be statistically significant according to the paired sample t-test ($p$-value $< 0.05$). Moreover, the poor performance of the TF-IDF as compared to the non contextual baseline (FTW) shows that the discriminatory power of a term does not necessarily correlate with the importance of that term towards defining the product intent of that query. Our competing approach VPCG \& VG gives worse ranking of the search results as compared to the baseline BM25F based retrieval algorithm, further demonstrating the advantage of using contextual knowledge to identify important terms from a search query, especially for the tail queries. Figure \ref{fig:mrr_length} shows how the MRR is affected with the increase in the query length. For all query lengths, CTW gives a better ranking of the search results than the baselines. However, for all the methods, the performance usually drops with the increase in query length, as noisy terms are expected to appear in longer queries.

The same trend is demonstrated by the other metrics too. The relative performance gain of the CTW over FTW, in terms of the $AP@nnz$ is $6.7\%$. We found the difference between the $AP@nnz$ for CTW and FTW to be statistically significant according to the paired sample t-test ($p$-value <  $0.01$). Even on $AP@1$, $AP@2$ and $AP@3$, CTW performs significantly better than the FTW, showing that the top $1, 2$ and $3$ predicted terms by the CTW model are better in explaining the query's product intent than the terms predicted by the FTW model. We provide further qualitative discussion regarding this in Section \ref{qualitative}.

\begin{table}[!t]
\small
\centering
  \caption{Results for the query refinement problem.}
  \begin{tabularx}{\columnwidth}{Xrrrr}
    \hline
Model    & $AP@nnz$  & $AP@1$  & $AP@2$  &$AP@3$\\ \hline
CQR{*} & \textbf{0.611} & \textbf{0.794} & \textbf{0.727} & \textbf{0.680}       \\
FQR & 0.591 & 0.753 & 0.691 & 0.657       \\ \hline
\end{tabularx}
  \label{tab:query_refinement}
   \begin{tablenotes}
        \small
        \item[*] The performance of the approach was found to be statistically significant ($p$-value $< 0.01$ using paired $t$-test).
   \end{tablenotes}
\end{table}

\subsubsection{Query refinement} Table \ref{tab:query_refinement} shows the performance statistics for the query refinement problem. Similar to the term-weighting problem, our proposed method contextual query refinement (CQR) beats the non-contextual baseline an all the metrics. The relative performance gain of CQR over the FQR, in terms of the $AP@nnz$ is $3.4\%$. We found the difference between the $AP@nnz$ for CQR and FQR to be statistically significant according to the paired sample t-test ($p$-value $< 0.01$). Similar to the term-weighting, CQR performs significantly better than the FQR even on the $AP@1$, $AP@2$ and $AP@3$ metrics, showing that the top $1, 2$ and $3$ predicted terms by the CQR model are better in explaining the query's product intent than the terms predicted by the FQR model. We provide further qualitative discussion regarding this in Section \ref{qualitative}.

\subsection{Qualitative evaluation}\label{qualitative}

\begin{table}

   \caption{Predicted term-weights for a some queries. The \textbf{bold} terms refer to the terms retained in the reformulated query.}\label{tab:weight_qualitative}
   \centering
   \subfloat[Initial query is ``battery night light with timer'' and the reformulated query is ``munchkin night light''.\label{tab:q1}]{
     \centering
     \begin{tabularx}{\columnwidth}{lRRRRR}\hline
       Model&battery&\textbf{night}&\textbf{light}&with&timer\\\hline
       CTW&0.73&0.80&1.00&0.51&0.29\\
       FTW&1.00&0.77&0.88&0.49&0.97\\
       TF-IDF&0.62&0.80&0.61&0.48&1.00\\
       VPCG \& VG&0.83&0.09&1.00&0.00&0.53\\\hline
     \end{tabularx}
   }\\
  \subfloat[Initial query is ``cars shaving kit'' and the reformulated query is ``kids shaving kit''.\label{tab:q2}]{
     \centering
     \begin{tabularx}{\columnwidth}{lRRR}\hline
      Model&cars&\textbf{shaving}&\textbf{kit}\\\hline
      CTW&0.58&1.00&0.77\\
      FTW&0.79&1.00&0.70\\
      TF-IDF&0.72&1.00&0.66\\
      VPCG \& VG&1.00&0.74&0.03\\\hline
     \end{tabularx}
  }\\
   \subfloat[Initial query is ``work boots steel toe breakers'' and the reformulated query is ``survivor work boots mens''.\label{tab:q3}]{
     \centering
     \begin{tabularx}{\columnwidth}{lRRRRR}\hline
       Model&\textbf{work}&\textbf{boots}&steel&toe&breakers\\\hline
       CTW&0.74&1.00&0.85&0.79&0.41\\
       FTW&0.51&0.76&0.65&0.72&1.00\\
       TF-IDF&0.62&0.51&0.58&0.69&1.00\\
       VPCG \& VG&0.01&1.00&0.56&0.01&0.12\\\hline
     \end{tabularx}
    }\\
  \subfloat[Initial query is ``12 piece gold flatware set'' and the reformulated query is ``flatware set for 12''.\label{tab:q4}]{
     \centering
     \begin{tabularx}{\columnwidth}{lRRRRR}\hline
      Model&\textbf{12}&piece&gold&\textbf{flatware}&\textbf{set}\\\hline
      CTW&0.71&0.63&0.55&1.00&0.89\\
      FTW&0.69&0.73&0.76&1.00&0.70\\
      TF-IDF&0.66&0.68&0.62&1.00&0.46\\
      VPCG \& VG&1.00&0.01&0.01&0.38&0.22\\\hline
     \end{tabularx}
  }\\
   \subfloat[Initial query is ``helmets for electric scooters for girls'' and the reformulated query is ``helmets''.\label{tab:q5}]{
     \centering
     \begin{tabularx}{\columnwidth}{lRRRRRR}\hline
       Model&\textbf{helmets}&for&electric&scooters&for&girls\\\hline
       CTW&1.00&0.90&0.82&0.69&0.79&0.66\\
       FTW&0.95&0.58&0.91&1.00&0.58&0.72\\
       TF-IDF&1.00&0.34&0.67&0.98&0.34&0.52\\
       VPCG \& VG&1.00&0.54&0.02&0.96&0.54&0.04\\\hline
     \end{tabularx}
   }\\
   \subfloat[Initial query is ``auto seat cover wonder woman'' and the reformulated query is ``auto seat cover''.\label{tab:q6}]{
     \centering
     \begin{tabularx}{\columnwidth}{lRRRRR}\hline
       Model&\textbf{auto}&\textbf{seat}&\textbf{cover}&wonder&woman\\\hline
       CTW&0.78&1.00&0.94&0.39&0.21\\
       FTW&0.52&0.85&0.59&0.91&1.00\\
       TF-IDF&0.92&0.71&0.74&1.00&0.73\\
       VPCG \& VG&0.19&0.10&0.07&1.00&0.16\\\hline
     \end{tabularx}
   }\\
\end{table}

\begin{table}

   \caption{Top 20 predicted terms for a few selected queries.}\label{tab:class_qualitative}
   \centering

   \subfloat[Initial query is ``orbit red garden hose water nozzle'' and the reformulated query is ``orbit hose nozzle''.\label{tab:c1}]{
     \centering
     \begin{tabularx}{\columnwidth}{lX}\hline
       Model&Top predicted terms\\\hline
       CQR& hose, nozzle, orbit, garden, water, red, high, wand, pressure, hoses, gilmour, flexible, rain, house, quick, nozzles, sprayer, valve, gutter, connector\\
       FQR& hose, water, garden, orbit, red, nozzle, better, gum, homes, timer, spray, bottle, pressure, washer, outdoor, rv, home, spearmint, adapter, black\\\hline
     \end{tabularx}
   }\\
   
  \subfloat[Initial query is ``little tikes balls for ball pit'' and the reformulated query is ``ball pit balls''.\label{tab:c2}]{
     \centering
     \begin{tabularx}{\columnwidth}{lX}\hline
      Model&Top predicted terms\\\hline
      CQR& ball, pit, balls, little, tikes, tykes, bounce, soccer, indoor, playground, pits, kids, plastic, sports, play, house, basketball, toys, put, set\\
      FQR& little, tikes, ball, balls, pit, fire, kids, tennis, set, toy, boss, golf, baby, toys, pony, play, girls, car, table, grill\\\hline
     \end{tabularx}
  }\\
   \subfloat[Initial query is ``physical therapy tools for ankle'' and the reformulated query is ``physical therapy tools''.\label{tab:c3}]{
     \centering
     \begin{tabularx}{\columnwidth}{lX}\hline
       Model&Top predicted terms\\\hline
       CQR& physical, therapy, tools, strap, ankle, tool, stretching, machine, braces, kids, yoga, wedge, pain, inversion, seller, foot, body, support, relief, hands\\
       FQR& therapy, ankle, socks, physical, tools, roller, foam, gift, weights, mens, black, body, kids, cards, set, cushion, womens, tool, oil, wedge\\\hline
     \end{tabularx}
    }\\

   \subfloat[Initial query is ``outdoor paint for house cream'' and the reformulated query is `exterior paint for house''.\label{tab:c4}]{
     \centering
     
     \begin{tabularx}{\columnwidth}{lX}\hline
       Model&Top predicted terms\\\hline
       CQR& paint, house, exterior, outdoor, spray, wood, white, metal, plastic, chalk, based, rust, kit, concrete, kits, gloss, gallon, cream, hide, color\\
       FQR&paint, outdoor, cream, house, spray, kids, white, set, ice, black, lights, christmas, light, acrylic, table, coffee, baby, dog, maxwell, face\\\hline
     \end{tabularx}
   }\\
   
  \subfloat[Initial query is ``kitchen shelving easy liner paper'' and the reformulated query is ``easy liner shelf liner''.\label{tab:c5}]{
     \centering
     \begin{tabularx}{\columnwidth}{lX}\hline
      Model&Top predicted terms\\\hline
      CQR& liner, easy, shelf, kitchen, paper, laminate, duck, x, shelving, oven, shelves, lining, liners, granite, roll, top, contact, plastic, rolls, peel\\
      FQR& paper, kitchen, shelving, liner, easy, shelf, wire, storage, shower, set, x, toilet, plastic, white, black, unit, wall, bags, towels, trash\\\hline
     \end{tabularx}
  }\\
\end{table}

\subsubsection{Term-weighting}
In order to visualize the weights produced by the CTW and compare them with the non-contextual baselines, we looked into a few selected search queries and the weights estimated for the terms in them by various methods. Table  \ref{tab:weight_qualitative} shows the predictions for some of the selected queries. The weights are normalized so that the maximum weight assigned to a term is one. Table \ref{tab:q1} shows an example of a query where the user reformulated it to capture the generalized product intent. The initial query is ``battery night light with timer'', for which the product intent is a \emph{night light} with some particular attributes. The product intent for the query was not satisfied, and the user reformulated the query to ``munchkin night light''. Without the contextual information, it would be difficult to estimate the product intent because of the presence of the terms like \emph{battery} and \emph{timer}. TF-IDF gives highest weight to the term \emph{timer}, because \emph{timer} is less frequent as compared to the other terms, but the term \emph{timer} will produce irrelevant results if it is not accompanied with the \emph{night light}. Similarly, FTW gives highest weight to the term \emph{battery} because \emph{battery} is a common term which defines the intent for many queries like ``battery operated fan'', ``battery charger'' etc. The term \emph{battery} is retained in a large number of reformulations, thus FTW gives it high weight. In a similar manner, VPCG \& VG gives higher weight to the term \emph{battery} than the term \emph{night} because \emph{battery} is a common term which defines the intent for many queries. In comparison, CTW is able to estimate the correct product type and gives highest weights to the terms \emph{night} and \emph{light}. Similarly, Table \ref{tab:q6} shows an example where the initial query is ``auto seat cover wonder woman'' and the reformulated query is ``auto seat cover''. CTW is able to estimate the correct type and give highest weight to the terms \emph{auto}, \emph{seat} and \emph{cover}, while other approaches failed to do so.

\subsubsection{Query refinement}
We looked into a few search queries and the top terms predicted for them by the CQR and compared them with the ones predicted by the FQR model. Table  \ref{tab:class_qualitative} shows the top 20 predicted terms for some of these queries. Table \ref{tab:c1} shows an example of a query and its reformulation. The initial query is ``orbit red garden hose water nozzle'', for which the product intent is a \emph{water spray nozzle} of brand \emph{orbit} and color \emph{red}. The product intent for the query was not satisfied, and the user reformulated the query to ``orbit hoze nozzle''. Without the contextual information, it is difficult to estimate the product intent because \emph{orbit} is also a popular chewing gum brand. FQR ends up predicting terms like \emph{gum} and \emph{spearmint} because of the presence of the term \emph{orbit}. In comparison, CQR is able to correctly estimate the correct product type \emph{water spray nozzle} and predicts only the terms relevant to it. Other examples also show a similar trend.

\section{Conclusion and future work}\label{conclusion}

In this paper, we leveraged the historical query reformulation logs of a major e-commerce retailer\footnote{walmart.com} to develop approaches that potentially can be used to decrease users' reformulation efforts, thus improving users' product search experience. We developed methods to identify important terms from the noisy terms in a search query and perform query refinement. Our methods improve upon the baselines by exploiting the context within the queries, that is, same product attribute can have different significance towards different product types. The proposed approaches perform better than the non-contextual baselines on multiple experiments, demonstrating the advantage of the proposed approaches.

To further verify the generalizability of our methods, it is important to use more test collections for evaluation. This constitutes an important future work. Apart from the query reformulations, another valuable source of information that can be used to identify the important terms in a query as well as to recommend new terms is the description of the engaged products. It would be an interesting research direction to combine these multiple sources of information to develop approaches to improve users' product search experience. To further evaluate our query refinement approach, it would be interesting to analyze the users' behaviour regarding how they modify their queries after going through the suggested terms. The code for this paper is available on GitHub \footnote{https://github.com/gurdaspuriya/query\_intent}.

\bibliographystyle{ACM-Reference-Format}
\bibliography{sample-bibliography}

\end{document}